\documentclass{ptephy_v1}

\preprintnumber{XXXX-XXXX} 

\newcommand{\be}{\begin{eqnarray}}
\newcommand{\ee}{\end{eqnarray}}
\newcommand{\nn}{\nonumber}

\def \Br{\mathop{\mbox{\normalfont Br}}\nolimits}
\def \dL{\mathcal{L}}
\def \GeV{\mathop{\mbox{\normalfont GeV}}\nolimits}
\def \TeV{\mathop{\mbox{\normalfont TeV}}\nolimits}
\def \Or{\mathcal{O}}

\def\half{{\textstyle{1 \over 2}}}

\newcommand{\yd}{{{Y^d_1}}}
\newcommand{\yydd}{{{Y^d_2}}}

\newcommand{\yu}{{{Y^u_1}}}
\newcommand{\yyuu}{{{Y^u_2}}}
\newcommand{\hggs}{\Phi_1}
\newcommand{\hggss}{\Phi_2}
\newcommand{\hgt}{\tilde{\Phi}_1}
\newcommand{\hgtt}{\tilde{\Phi}_2}

\begin{document}

\title{Probing flavor parameters in the scalar sector  and new bounds for the fermion sector}


\author{R. Gait\'an}
\author{J.H. Montes~de~Oca}
\author{J.A. Orduz-Ducuara}
\affil{Departamento de F\'isica, FES Cuautitl\'an,\\ 
Universidad Nacional Aut\'onoma de M\'exico\\
Estado de M\'exico, 54770, M\'exico \email{rgaitan@unam.mx}}

%
%

\begin{abstract}
In this paper we study the 
Flavor-Changing mediated by 
Higgs boson within Two-Higgs Doublet Model-III context. 
We explore the parameter space
and, considering recent results, find 
new limits for the model parameters.  
We also  
obtain the total Higgs decay width 
and the branching ratios for 
different channels taking limits for 
the $t\to cV$ and $b\to s\gamma$ 
processes.	
Considering different constraints, we 
estimate the branching ratio for the 
$h\to t^*_{}c$ in the model as well 
as the bounds for the $b \to s\gamma,$ 
$h \to \mu\tau$ and $h\to \gamma Z.$ 
Considering the quark top decays to 
$W\bar{b},$ we obtained 
${\Br}(h \to Wc\bar{b})\sim10^{-3}.$
\end{abstract}

\subjectindex{B40, B46}

%

\maketitle


\section{\label{sec:Intro} Introduction}

The Standard Model (SM) describes the universe at scale $\Or(10^{2} \GeV).$ 
Different experiments 
have examined the SM observables 
which have good agreement 
with the theoretical predictions.
Currently LHC explores the nature at energy 
scale of order  $\TeV'$s  and the SM works 
very well to analyze the structure 
of the matter. 
However there are different questions about the 
unknown phenomena: 
the matter-antimatter asymmetry, 
the CP violation and 
flavor-changing neutral currents (FCNC) mediated by gauge 
and scalar bosons. In this paper
we shall explore the last one where 
the neutral scalar boson can change 
the fermion flavor; we focus on the 
$h\to VV'$ processes where the 
$h$ is the lightest scalar and 
$V, V' = Z, \gamma, g.$ 
 
We shall discuss the Flavor-Changing Neutral 
Scalar Interactions (FCNSI) inside the Two-Higgs 
Doublet Model type III (THDM-III) context  considering different 
scenarios for the parameters of the model. 
Previous studies have considered 
different versions for this THDM-III 
\cite{Arroyo:2013tna, HernandezSanchez:2012eg, DiazCruz:2004pj, Kopp:2014rva, Harnik:2012pb},  
we shall focus on the Flavor-Changing (FC) mediated 
by scalars and then study Higgs decay processes. 
The FC context mediated by scalar 
bosons where the rare top decays have been 
studied, in particular in the THDM, 
the literature have reported the following branching ratios (${\Br}$) \cite{TheATLAScollaboration:2013nia, 
Eilam:1990zc, Atwood:1996vj}:
$ \Br{(t \to uh)} = 5.5\times 10^{-6},$ 
$ \Br{(t\to ch)} =1.5\times 10^{-3},$  
$ \Br{(t\to u \gamma)}  \sim10^{-6},$ 
$ \Br{(t\to c\gamma)}  \sim10^{-6},$ 
$ \Br{(t\to uZ)}  \sim10^{-7},$ 
$ \Br{(t\to cZ)}  \sim10^{-7}.$

There are different theoretical and experimental 
motivations to explore 
FC. Theoretically, for instance, 
to constrain the parameter space, 
and then obtain 
bounds over the parameters of the model. 
If we introduce new fermions,  
is possible to obtain the FC, at tree-level, 
when the new fermions and the SM fermions are mixed  
\cite{La:2013gga, Langacker:2000ju, Ponce:1992je}.  
On the other side, if we introduce a new gauge group, and 
consider that the SM fermion charges related to extra gauge 
group are non-universal family, it could generate the 
FC as was shown in refs. \cite{Barger:2009qs, Cleaver:1997jb, Masip:1999mk, DiazCruz:2012xa, Cotti:2002zq}.
The motivation to explore the FC is to test the SM 
and its behavior faced on New Physics scenarios.

From an experimental point of view, the uncertainties 
in the results motivate the flavor physics. Recently, ATLAS and CMS have published 
results for FC processes: 
$l_i \to l_j ~\gamma $ and $l_i \to l_j ~ h$ 
\cite{Aad:2015gha, Khachatryan:2015kon, Agashe:2014kda,TheATLAScollaboration:2013nia, CMS-PAS-TOP-14-020}. Besides, other 
collaborations have reported limits for 
the FC in the lepton sector:  
$\Br(\tau^- \to \mu^- \mu^+ \mu^-) = 2.1\times 10^{-8}$ 
\cite{Agashe:2014kda}, 
${\Br}(\tau^- \to \mu^-\pi^+ \pi^-)=$ 
$10^{-7}$ (Belle),
$10^{-6}$ (BaBar), 
$10^{-5}$ (CLEO)
\cite{Amhis:2012bh}. 
From LHCb 
results on $B^+ \to K^+ \mu^+ \mu- (e^+e^-)$ 
process have motivated the studies on the non-universal 
family couplings \cite{LHCP-2014, Lees:2012xj}, 
the orthogonality of the 
CKM matrix \cite{Agashe:2014kda, Quang:1998yw} and the universality violation \cite{Fajfer:2012jt}.

Our paper is organized as follows:
In section \ref{sec:Metho}, 
we discuss the model 
and methods to consider the 
FC mediated by scalar bosons. 
We show the ways to obtain the FC at 
loop level in the $h\to VV'$ process
mediated by fermions. 
We implemented SARAH 
\cite{Staub:2015kfa} for THDM type III, 
the amplitudes are obtained by 
FeynArts \cite{Hahn:2000kx} and the 
Passarino-Veltman functions are evaluated 
by LoopTools \cite{Hahn:1998yk}.
In section \ref{sec:Resul} we present
the results considering our parametrization 
for the $\Br$, as well as decay width. 
We introduce bounds coming from 
meson processes which are taken 
from ref. \cite{ElKhadra:2002wp}
Section \ref{sec:Discu-Concl} contains a 
discussion of our results and the conclusions.

\section{\label{sec:Metho} Model and Methods}

We shall consider the THDM-III  where the most 
general potential{\footnote
{The potential in different parametrizations 
can be found in 
\cite{Gunion:1989we, HernandezSanchez:2012eg}}} is  
\cite{Gunion:2002zf}
\be
V(\Phi_1 \Phi_2) &=&
m_{11}^2\Phi_1^\dagger\Phi_1+m_{22}^2\Phi_2^\dagger\Phi_2
-[m_{12}^2\Phi_1^\dagger\Phi_2+{\rm h.c.}]
\nn\\
&&
+\half\lambda_1(\Phi_1^\dagger\Phi_1)^2
+\half\lambda_2(\Phi_2^\dagger\Phi_2)^2
\nn\\
&&
+\lambda_3(\Phi_1^\dagger\Phi_1)(\Phi_2^\dagger\Phi_2)
+\lambda_4(\Phi_1^\dagger\Phi_2)(\Phi_2^\dagger\Phi_1)
\nn\\
&& +\Big\{\half\lambda_5(\Phi_1^\dagger\Phi_2)^2
+\big[\lambda_6(\Phi_1^\dagger\Phi_1)
\nn\\
&&
+\lambda_7(\Phi_2^\dagger\Phi_2)\big]
\Phi_1^\dagger\Phi_2+{\rm h.c.}\Big\}\,. \nn
\ee
where $m_{ii}^{2}, $ $\lambda_i$ are reals, 
and $\Phi_{1,2}$ denote the complex doublet 
 scalar fields. The scalar masses spectrum can be 
 found in detail in ref. \cite{Branco:2011iw}. The perturbativity condition for the the validity of a tree approximation in the interactions of the SM-like Higgs boson is imposed by requiring that the couplings $|\lambda_i| \leqslant 4 \pi$
 
The interaction and physical states for the scalars are related through the mixing parameters $\alpha$ and $\beta${\footnote{the $\tan\beta$ is more convenient to use for  the analysis. We denote $\tan\beta=t_\beta.$}}. We assume the alignment limit \cite{Wang:2016rvz}, which states a relation between $\alpha$ and $\beta,$ such as $\beta- \alpha=\pi/2-\delta$ for $\delta <<1,$ corresponding to the decoupling limit \cite{Gunion:2002zf}. In particular, we take $\delta = 1\times10^{-2}$,  which means the coupling of the lightest Higgs to fermions and gauge bosons is SM-like.

In a general way, the Yukawa sector for the THDM-III 
is given by
\be\label{eq:lgrngn-THDMIII}
 {\dL}^{THDM-III}_{YS} &=& 
 {\yu}\overline{Q}_{L}^0 {\hgt} u_{R}^0  +
{{\yyuu}} \overline{Q}_{L}^0 {\hgtt} u_{R}^ 0
\nn\\&&
 + {{\yd}} \overline{Q}_{L}^0 {\hggs} d_{R}^0 
  +
{{\yydd}} \overline{Q}_{L}^0 {\hggss} d_{R}^0 + h.c.
\ee
with 
$
Q_L^0 = \left(
 \begin{matrix}
u_L &
d_L 
 \end{matrix}
\right)^{T}, 
{\Phi_{1,2}^{}} = \left(
  \begin{matrix}
\phi_{1,2}^\pm &
\phi_{1,2}^{0} 
  \end{matrix}
\right)^{\dagger}, 
$ 
and 
$
\tilde{\Phi}_j = i \sigma_2\Phi_j^*.$
$Y_i$ are the Yukawa matrices, 
$\phi^0_{i}$ is the 
neutral Higgs eigenstate, 
and $\phi^{\pm}$ contains the 
charged pseudo-goldstone boson. 
Analogously, the Yukawa interaction 
can be written for leptons.

For the neutral scalar fields the Yukawa Lagrangian  
in the mass eigenstates is 
given by
\be\label{eq:Lgrnn-gral-THDM-III}
{\dL}^{THDM-III}_{n} &=& 
\frac{g}{2}
\left(\frac{m_{i}}{m_{W}}\right)\bar{d}_i^{}
\left[
\frac{\cos\alpha}{\cos\beta}\delta_{ij}+\frac{\sqrt{2}\sin(\alpha-\beta)}{g\cos\beta}\left(\frac{m_{W}}{m_{i}}\right)\left(\tilde{Y}_{2}^{d}\right)_{ij}
\right]
d_{j}^{}H^{0} \nonumber\\
 & + & \frac{g}{2}
 \left(\frac{m_{j}^{}}{m_{W}}\right)\bar{d}_{i}^{}
 \left[
 -\frac{\sin\alpha}{\cos\beta}\delta_{ij}^{}+
 \frac{\sqrt{2}\cos(\alpha-\beta)}{g\cos\beta}
 \left(\frac{m_{W}}{m_{i}}\right)
 \left(\tilde{Y}_{2}^{d}\right)_{ij}^{}
 \right] d_{j}^{}h^{0} \nn\\
 & + & \frac{ig}{2}
 \left(\frac{m_{i}}{m_{W}}\right)\bar{d}_{i}^{}
 \left[
 -\tan\beta\delta_{ij}^{}+\frac{\sqrt{2}}{g\cos\beta}
 \left(\frac{m_{W}}{m_{i}}\right)
 \left(\tilde{Y}_{2}^{d}\right)_{ij}^{}
 \right]\gamma^{5}d_{j}^{}A^{0} \nonumber \\
 & + & \frac{g}{2}
 \left(\frac{m_{i}}{m_{W}}\right)\bar{u}_{i}^{}
 \left[
 \frac{\sin\alpha}{\sin\beta}\delta_{ij}^{}+\frac{\sqrt{2}
 \sin(\alpha-\beta)}{g\sin\beta}\left(\frac{m_{W}}{m_{i}}\right)\left(\tilde{Y}_{2}^{u}\right)_{ij}^{}
 \right]u_{j}^{}H^{0} \nn\\
 & + & \frac{g}{2}
 \left(\frac{m_{u}}{m_{W}}\right)\bar{u}_{i}^{}
 \left[
 -\frac{\cos\alpha}{\sin\beta}\delta_{ij}^{}+
 \frac{\sqrt{2}\cos(\alpha-\beta)}{g\sin\beta}
 \left(\frac{m_{W}}{m_{i}}\right)
 \left(\tilde{Y}_{2}^{u}\right)_{ij}^{}
 \right]u_{j}^{}h^{0} \nn \\
  &+ & \frac{ig}{2}
  \left(\frac{m_{u}}{m_{W}}\right)\bar{u}_{i}^{}
  \left[
  -\cot\beta\delta_{ij}^{} + 
  \frac{\sqrt{2}}{g\sin\beta}
  \left(\frac{m_{W}}{m_{i}}\right)
  \left(\tilde{Y}_{2}^{u}\right)_{ij}^{}
  \right]\gamma^{5}u_{j}^{}A^{0}.
\ee
The complete Yukawa Lagrangian can be revised, for instance,  in ref. \cite{Arroyo:2013tna}.

\subsection{FC by scalar sector}

Next we shall describe a way to parametrize 
the FCNSI 
\cite{Diaz-Cruz:2014aga, DiazCruz:2004pj}.
In general each of the Yukawa matrices are 
non-diagonal; however their linear combination,
which is the mass matrix, is diagonal.
The FC is given when 
the diagonalization of the fermion 
mass matrices does not imply 
the diagonalization of the Yukawa 
couplings.

In this sector the interaction between 
fermions and scalar is given by the 
Yukawa couplings ($Y_f^{}$).
In the SM the couplings is given by 
\be
g_{hff}^{SM} = Y_{SM}^{f} = \sqrt{2}\frac{m_f^{}}{v} 
\nn
\ee
where $m_f^{}$ is fermion mass and $v$ is the vaccum expectation 
value (VEV). 

For the THDM-III, 
after the spontaneous symmetry breaking 
the mass matrix is 
\be\label{eq:mf-dgnlsd-frmns}
M_f^{}=\frac{1}{\sqrt{2}} 
\Big(v_1^{}Y_1^{f}  + v_2^{}Y_2^{f}\Big)
\ee 
where $v_1^{}$ and $v_2^{}$ 
are VEV's associated to the $\phi_1^{0}$ 
and $\phi_2^{0}$, respectively, and are related by 
$t_\beta^{} =\frac{v_2^{}}{v_{1}^{}}.$
$Y^{u,d}_{}$ is the Yukawa couplings 
that could be complex. 
In a general form, the 
eq. \eqref{eq:mf-dgnlsd-frmns} 
is non-diagonal; it can be diagonalized
through biunitary transformation $U_{L,R}:$
\be
\big(U^{f}_{L}\big)^{\dagger} 
M_f^{}
\big(U^{f}_{R}\big)^{}
=
{\widetilde{M}}_f^{}\nn
\ee
Note that the  
widetilde over the quantities means 
the physics basis.

In order to reduce the free parameters, 
one can rewrite the 
eq.\eqref{eq:mf-dgnlsd-frmns} as
\be\label{eq:Y-rewrite}
{\widetilde{Y}}_1^{f} &=& 
\sqrt{2} 
\frac{M_f^{}}{v_{}^{}\cos\beta} - 
{\widetilde{Y}}_2^{f} ~t_\beta^{}, \nn
\ee 
and substitutes in 
eq. \eqref{eq:lgrngn-THDMIII}.

Then it is possible to have mixing 
between the fermion flavor at tree level. 
On going, we suppress the tilde and index in the 
the Yukawa matrix.

\subsection{\label{sec:FCNSI-LL} FCNSI at loop-level}

The FC processes at loop-level have some restrictions.
At this level, the effects must be much 
smaller than the effects at tree level.  
The FCNSI is  an 
interesting topic for the high precision 
studies on the scalar sector 
because it could give a signal of 
new physics.
%
%
\begin{figure}[!h]
\centering\includegraphics[width=2.5in]{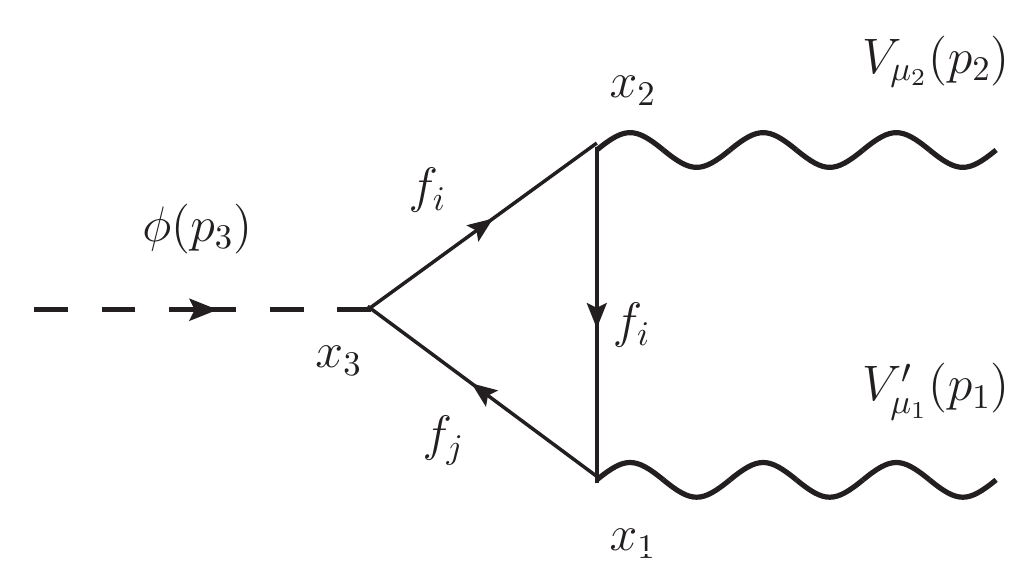}
\caption{The Feynman diagrams for the  $h \to VV'$. $x_1^{}$ and $x_2^{}$ vertices gets highly suppressed if the $V=\gamma$  while the $x_3^{}$ vertex contains the FC.}
\label{fig:phi-VV-NFV}
\end{figure}
%
%

We analyze the $h \to VV'$ process shown in 
fig. \ref{fig:phi-VV-NFV}, which is 
at higher level of precision 
where the remarked vertices $(x_i^{} )$ imply 
loop correction. 
{\bf{$x_{1,2}$-vertices:}} neutral vector  
boson and two fermions with 
FCNC.
{\bf{$x_{3}$-vertex:}} scalar boson and two fermions 
with FCNC mediated by the  
scalar bosons, this vertex contains 
FCNSI at tree-level. 
%
%
\begin{figure}[!h]
\centering\includegraphics[width=2.5in]{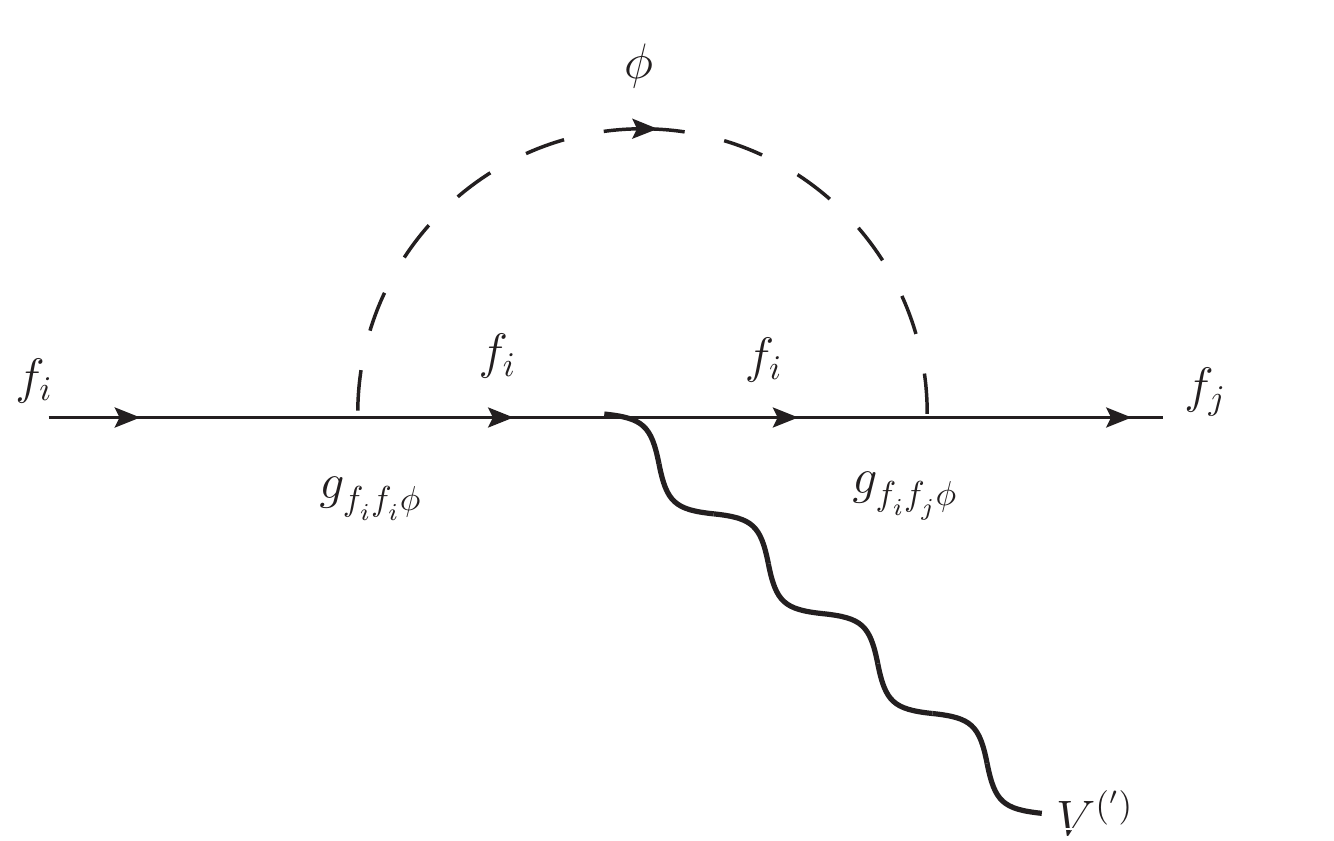}
\caption{The $f_i f_j V$ at one-loop level mediated by the scalar boson.}
\label{fig:2-loops-H-AA-fermions}
\end{figure}
%
%

In general, the couplings 
are given by $
g_{ijh}^{} = 
Y^{*}_{ij} 
P_L^{} + Y^{}_{ij} P_R^{},$ which 
represents the 
Higgs and pair of fermions 
vertex, $P_{R, L}^{}$ 
are projectors and $i,j$ are the 
generation of fermions. 
Constraints for the 
parameters come from 
experimental data, e.g.:
$
\sqrt{\big|Y_{\mu \tau}\big|^{2} 
+ 
\big|Y_{\tau \mu}\big|^{2}
} <3.6\times10^{-3}_{}
$\cite{Khachatryan:2015kon}.
Other constraints on the FC Yukawa 
couplings $\big(Y_{ij}^{}\big)$ are shown in 
ref. \cite{Alte:2016yuw}.

\section{\label{sec:Resul} Phenomenological results}

In this section we show  
phenomenological results. First part shows,  
a exploration on the parameter space which 
contains the 
phenomenological results for the parameter 
space: 
$\chi_{ij}^{u}-\chi_{ij}^{d}.$ 
In second part, we show the total  decay width for the 
Higgs and the ${\Br}$'s  for 
processes $h\to b\bar{b}, l\bar{l}, gg, 
\gamma\gamma$ and $Z\gamma;$ 
and also the allowed 
regions for different values of $t_\beta^{}.$ 
Moreover we take the recent  
results $t\to cZ, cg; 
b\to s \gamma$ and $h\to \gamma Z$ and 
explore the parameter space 
whithin the THDM-III context 
\cite{ElKhadra:2002wp}. 
\subsection{Constraints for the 
flavor parameters mediated by scalar boson}

In order to explore the Yukawa couplings 
we use the 
Cheng-Sher parametrization 
\cite{Cheng:1987rs}{\footnote{
This relation has been used in different papers
\cite{Arroyo:2013tna, GomezBock:2005hc, Diaz-Cruz:2014aga}.
Different analysis on the 
Yukawa couplings have explored the 
FC some results can be found in 
ref. \cite{Gaitan:2015hga, AguilarSaavedra:2004wm}.
}}  
\be\label{eq:nztsCS}
{Y}_{ij}^{f} = 
\sqrt{2} \frac{\sqrt{m_i m_j}}{v}{\chi}_{ij}^{f} 
\ee
where $m_i$ and $m_j$ are 
the fermion masses, and 
${\chi}_{ij}^{f}$ 
are free dimensionless parameters which 
shall probe the 
flavor-changing mediated by scalar bosons. 
This kind of parametrization is interesting
to probe the parameter space, and 
explore the flavor parameters in the 2HDM-III. 
An completed review on THDM can be found in 
\cite{Branco:2011iw}; our paper uses the current 
constraints and explores the some observables 
versus the flavor parameters.

For our research, we consider 
 the $\chi$-parameters 
dependent on type quark because 
there is not reason to assume that 
all those are equal as was mentioned 
in ref.\cite{Atwood:1996vj}. Besides we 
will label them as $\chi_{ij}^{u}$ and 
$\chi_{ij}^{d}.$   
For this work we will include the 
contribution coming from the 
heavy fermions. 

\begin{figure}[!h]
\centering
\includegraphics[width=2.5in]{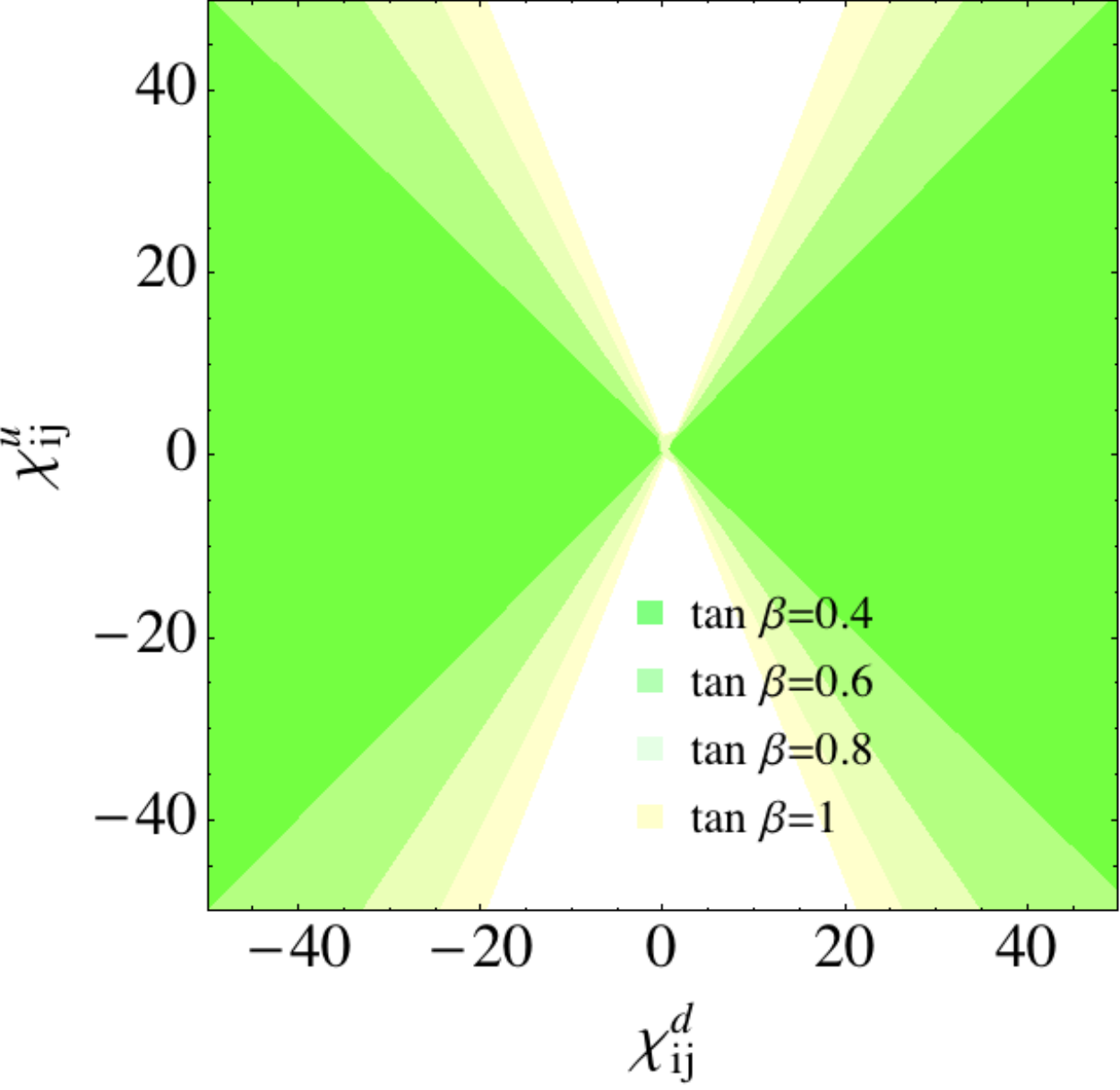}
\caption{Parameter space for the 
$\Gamma_t^{}$ (total decay width) for the Higgs boson,  we consider $m_h =125 \GeV$,  ${\Br}(h\to \gamma\gamma) = 2.27\times 10^{-3}$, $m_H= 300 \GeV$ and 
$m_{H^\pm} = 350 \GeV$.}
\label{fig:chiuu-chidd-THDM}
\end{figure}

We consider the experimental bound for the 
${\Br} (h\to \gamma \gamma)$ and explore 
the parameters of the THDM-III. Fig. 
\ref{fig:chiuu-chidd-THDM} shows the 
allowed regions for the parameter space 
taking different values for the $t_\beta^{}$
parameter.
%
%
%
%

We assume the universality for parameters $\chi_{ij}^{u,d}$, which means the same order for all $\chi_{ij}^{u,d}$. The hierarchy for Yukawa couplings arise from the fermion masses, as shows the Cheng-Sher ansatz in eq. \ref{eq:nztsCS}. The up-quark Yukawa couplings associated with $-40 \leqslant \chi_{ij}^{u,d}\leqslant 40$ are:
\begin{eqnarray}
|Y_{11}^{u}| \leqslant 5.2 \times 10^{-4},  \nonumber \\ 
|Y_{21}^{u}| \leqslant 0.012,  \nonumber \\
|Y_{22}^{u}| \leqslant 0.27,  \nonumber \\
|Y_{31}^{u}| \leqslant 0.14, \nonumber \\
|Y_{32}^{u}| \leqslant 3.3, \nonumber \\
|Y_{33}^{u}| \leqslant 39.6,
\end{eqnarray}
meanwhile for down-quark are:
\begin{eqnarray}
|Y_{11}^{d}| \leqslant 9.1 \times 10^{-4},  \nonumber \\ 
|Y_{21}^{d}| \leqslant 0.0043,  \nonumber \\ 
|Y_{22}^{d}| \leqslant 0.02,  \nonumber \\ 
|Y_{31}^{d}| \leqslant 0.031,  \nonumber \\ 
|Y_{32}^{d}| \leqslant 0.14,  \nonumber \\ 
|Y_{33}^{d}| \leqslant 1.15.
\end{eqnarray}

We note that the interactions between neutral scalars and quarks have proportional terms to $\cos(\alpha - \beta)$ or $\sin(\alpha - \beta)$. If the alignment limit is considered then $\cos(\alpha - \beta) \approx 0$ and $\sin(\alpha - \beta) \approx 1$. In particular, the Feynman rules for lightest neutral scalar and quarks contain the Yukawa couplings with factor  $\cos(\alpha - \beta)$ and additional terms like $\frac{\cos(\alpha)}{\sin(\beta)}$ for up quarks or  $\frac{\sin(\alpha)}{\cos(\beta)}$ for down quarks. This linear combinations appear in the amplitude for $h\rightarrow \gamma \gamma$ decay in the following form: 
\begin{equation}
|M|^2 \sim K_1 |Y_{ij}^{u} \cos(\alpha - \beta) - \frac{\cos(\alpha)}{\sin(\beta)}|^2 + K_2 |Y_{ij}^{d} \cos(\alpha - \beta) - \frac{\sin(\alpha)}{\cos(\beta)}|^2,
\end{equation}
where $ K_1 $ and  $K_2$ contain the Passarino-Veltman integrals involved in the decay at one loop. In Fig.4 the linear behavior arise from the smallness in the alignment limit and the inverse behavior is attribute to the inversions in the mixing parameters for the quarks Yukawa interactions.

\subsection{Branching ratios for the lightest neutral scalar}

Within the THDM-III context is feasible the 
FC exploration and the structure of the 
Yukawa couplings. In this work, we consider 
universality between fermions of different 
generations. We have inspired in the Yukawa 
couplings structure coming from textures
(see eq. \eqref{eq:nztsCS}), 
which are dependent on the one parameter related 
to the quark type 
\cite{GomezBock:2009xz, Arroyo:2013tna}.

\begin{figure}[!h]
\centering
\includegraphics[scale=0.6]{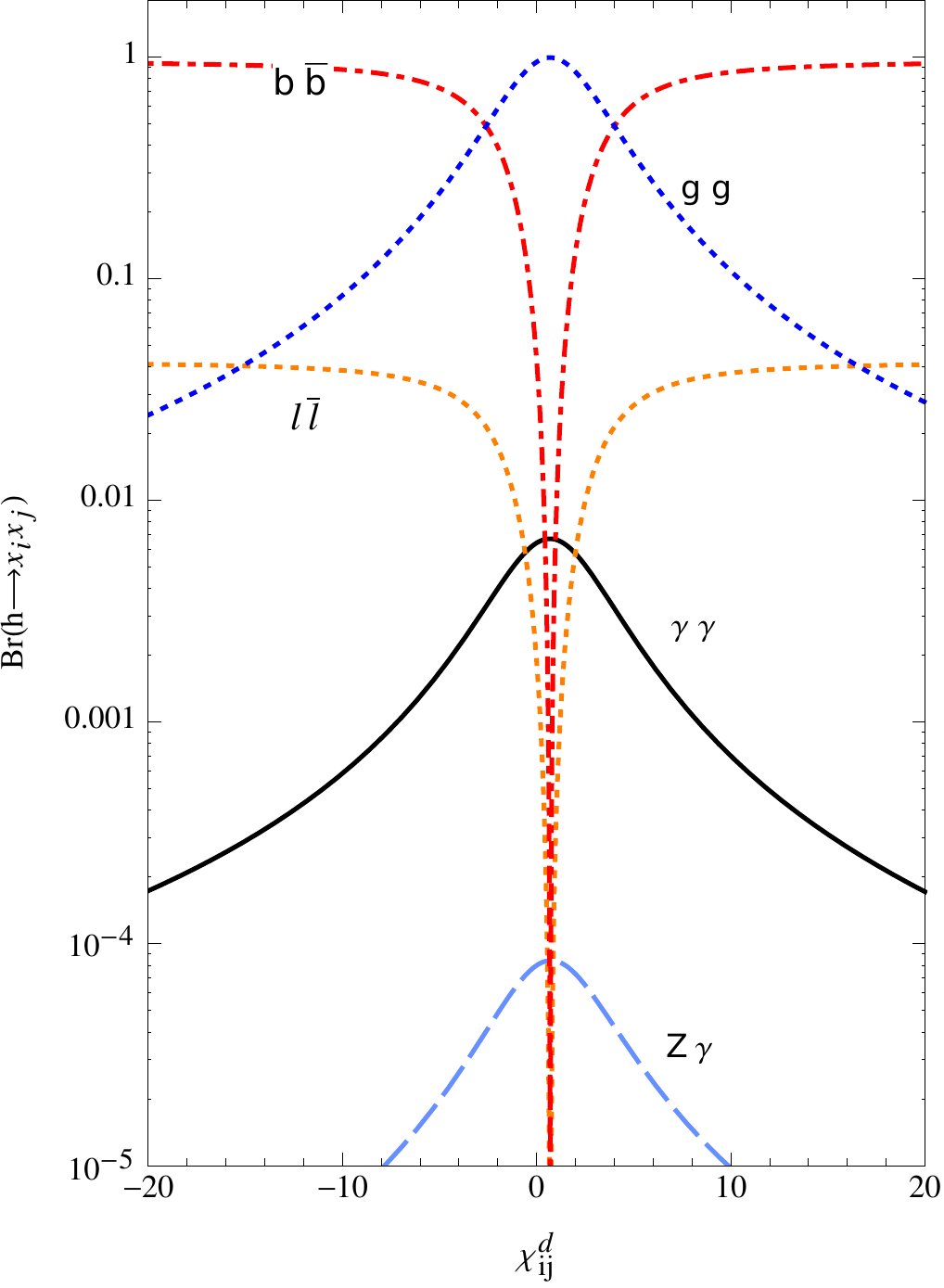}
\includegraphics[scale=0.6]{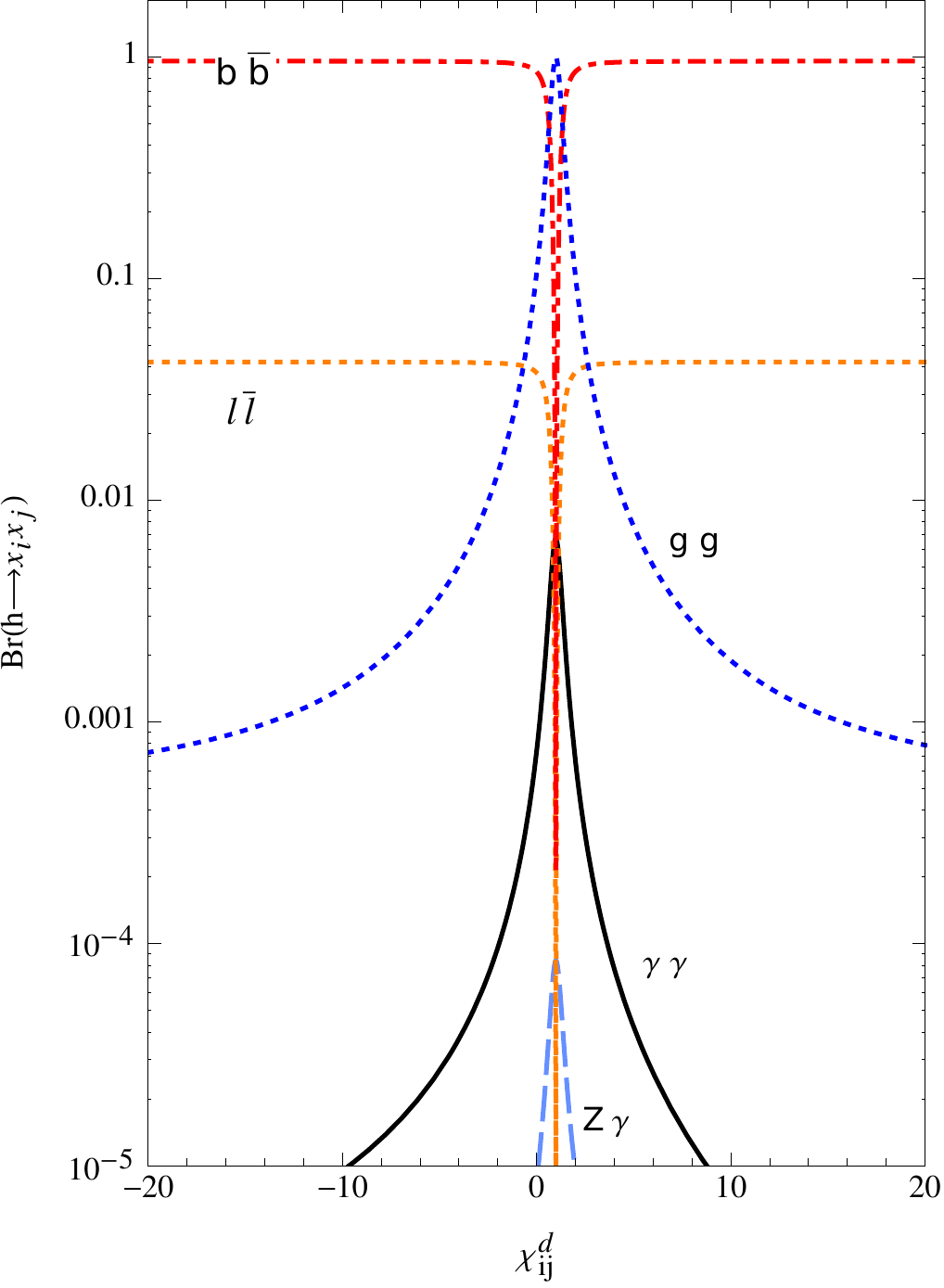}
\caption{$\Br$'s for different channels versus the flavor parameters for the d-type quark. We consider: a) (left-side) $t_\beta^{} = 1$ and b) (right-side) $t_\beta^{} = 10$, andfor both cases $\chi_{ij}^{u} = 12$.}
\label{fig:Br0101-Br011}
\end{figure}

As it is know the FCNSI are suppressed,  
e.g.  the ${\Br}(h\to \mu \tau) \lesssim10^{-2}$
\cite{Khachatryan:2015kon}. More limits have been 
explored from the top physics \cite{Eilam:1990zc, Gaitan:2015hga}. Our results are shown in fig. \ref{fig:Br0101-Br011} considering two scenarios for $\tan \beta=1,10.$
We show fermion and boson decays, where we 
use the labeled style line with the final states to represent 
our results and the thicker style line are 
the experimental results. 
This figure shows excluded range for the different 
mode decays versus the $\chi_{ij}^{d}.$ 

We explore the Higgs decays to pair bosons 
at one-loop level. 
To explore those channels, we used the 
Cheng-Sher parametrization for the fermions 
and consider the FCNSI.

%
%
%
%
%

Fig.  \ref{fig:Br0101-Br011} shows 
modes of the $h$ decay versus the $\chi_{dd'}^{},$  we found some 
enhancement regions for these channel with 
neutral boson in the final states.
The 
${\Br}$'s for different processes inside the THDM-III context and their observed results \cite{Dittmaier:2012vm, Barger:2012hv}. We show different scenarios for the 
parameters $\chi_{ij}^{u}, \chi_{ij}^{d}$ and 
$t_\beta^{}.$

\subsection{\label{sec:LHC-cnstrnts} Constraints from the recent experimental results}

We consider the recent experimental 
results coming from 
\cite{Olive:2016xmw} for processes $t\to c~V$
that could give evidence for the new physics.
This kind of processes are stu\-di\-ed in 
THDM-I,II and III context \cite{Atwood:1996vj, 
Gaitan:2015hga} and 
their results are constrained, however those data 
can get enhancement as the same manner as 
there is an enhancement in the Yukawa coupling 
exploration. 

\begin{figure*}[!h]
\centerline{
(a)
\includegraphics[width=2.5in]{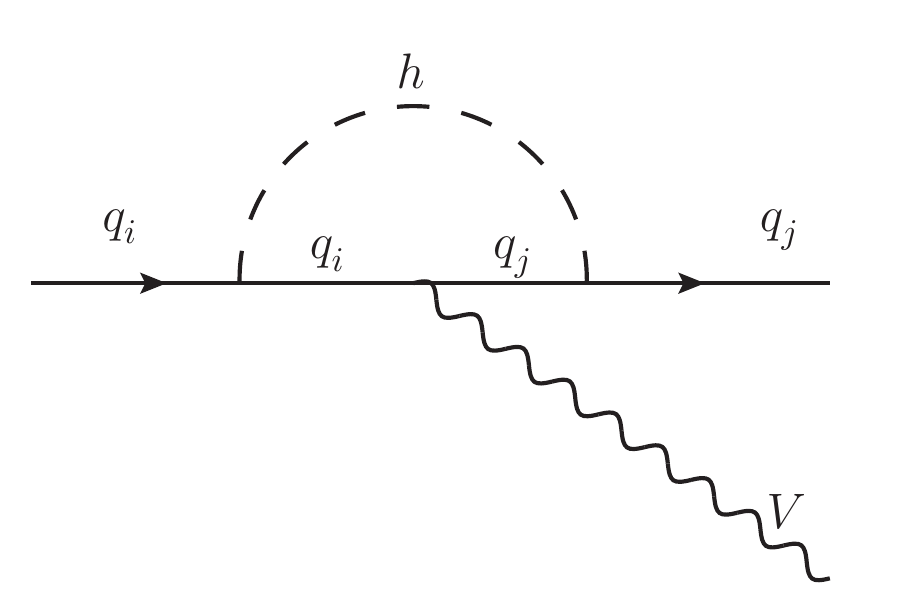}
\label{fig:fig1_li-lj}
\hfil
(b)
\includegraphics[width=2.5in]{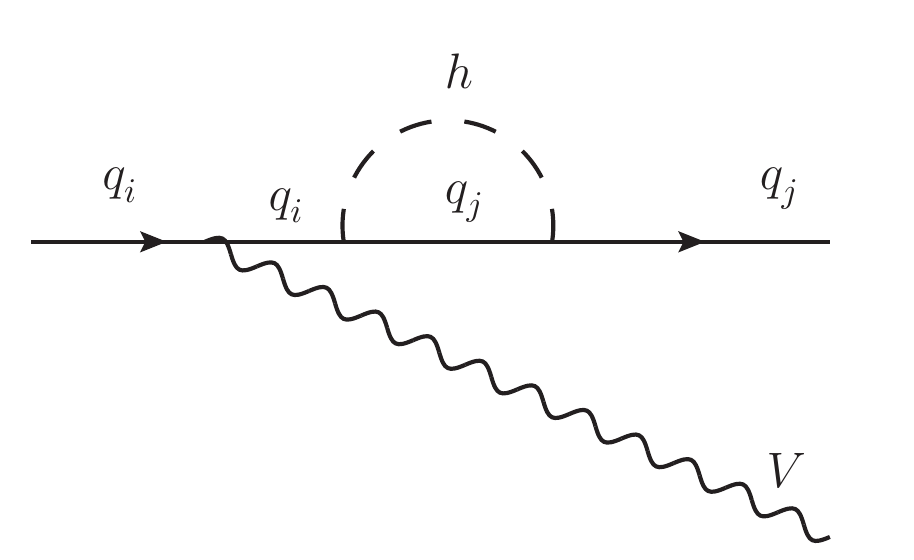}%
\label{fig:fig2_li-lj}
}
\hfil
(c)
\includegraphics[width=2.5in]{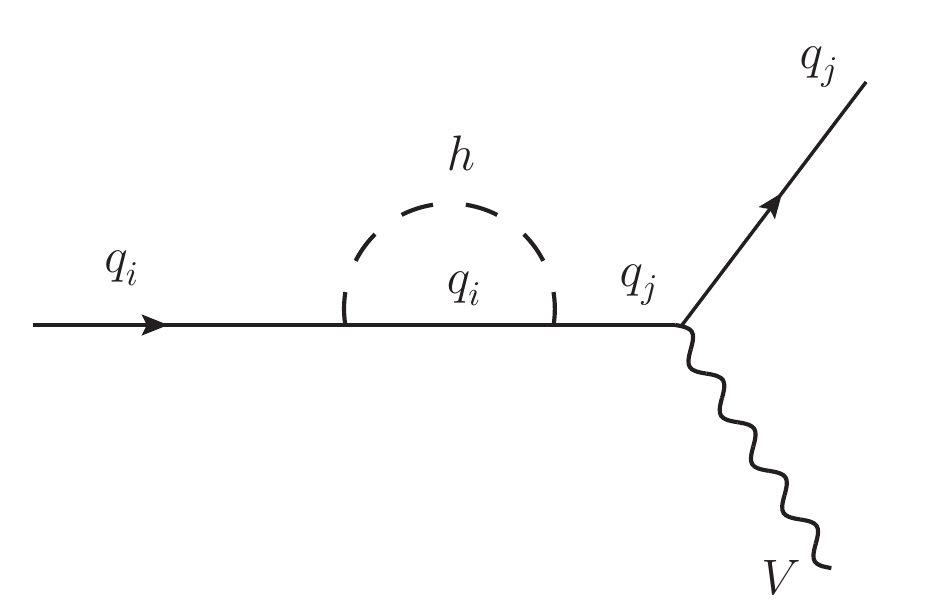}%
\label{fig:fig3_li-lj}
\caption{Feynman diagrams for the fermion decays with $q_i^{}$ u- and d-type, and $V = \gamma, Z, g.$ We considered the contributions  to the FC mediated by scalar bosons. Represented contributions are needed to avoid divergences.}
\label{fig:FD-t-cV}
\end{figure*}

We use the Feynman diagrams shown in 
fig. \ref{fig:FD-t-cV}.
In order to probe the beyond standard model 
physics and get strong constraints over the 
flavor parameters, we introduce 
$B \to X_s \gamma$ processes considering 
$\Gamma (B \to X_s^{}\gamma) \simeq 
\Gamma (b \to s^{}_{}\gamma),$ since 
the non-perturbative effects are small 
\cite{ElKhadra:2002wp}. The experimental limit for $Br (B \to X_s^{}\gamma)$, reported by \cite{Agashe:2014kda},  is used to constraint the numerical analysis.

%
%
%

%
%

Processes shown in fig. \ref{fig:FD-t-cV} 
constrain Flavor parameters. These processes allow to perform 
precision studies on the top decay 
\cite{AguilarSaavedra:2002ns}. 
Besides it is possible 
to analyze the FC at loop-level considering 
a neutral scalar.
 
\begin{figure*}[!h]
\centerline{
(a)
\includegraphics[width=2.5in]{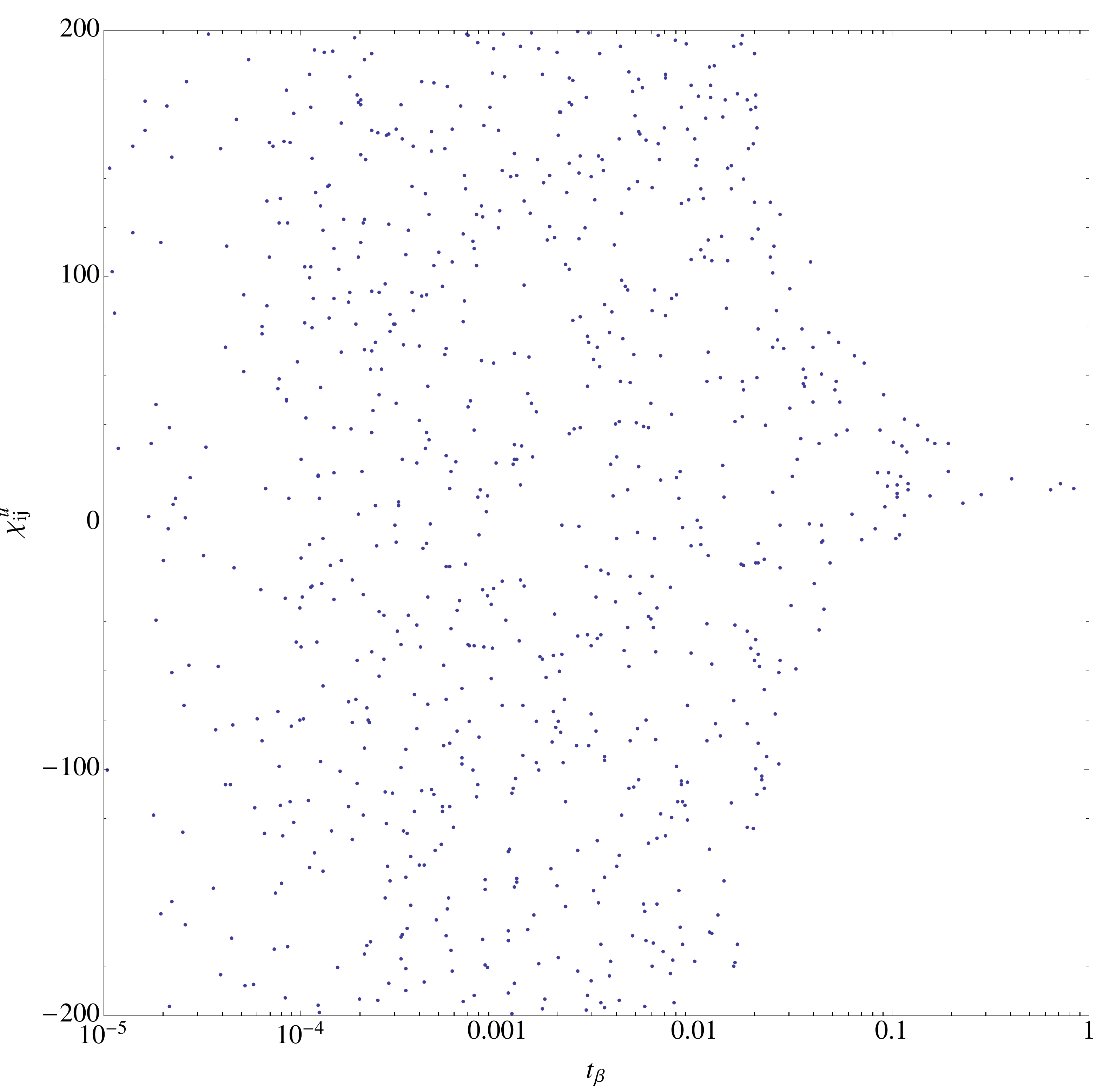}
\hfil
(b)
\includegraphics[width=2.5in]{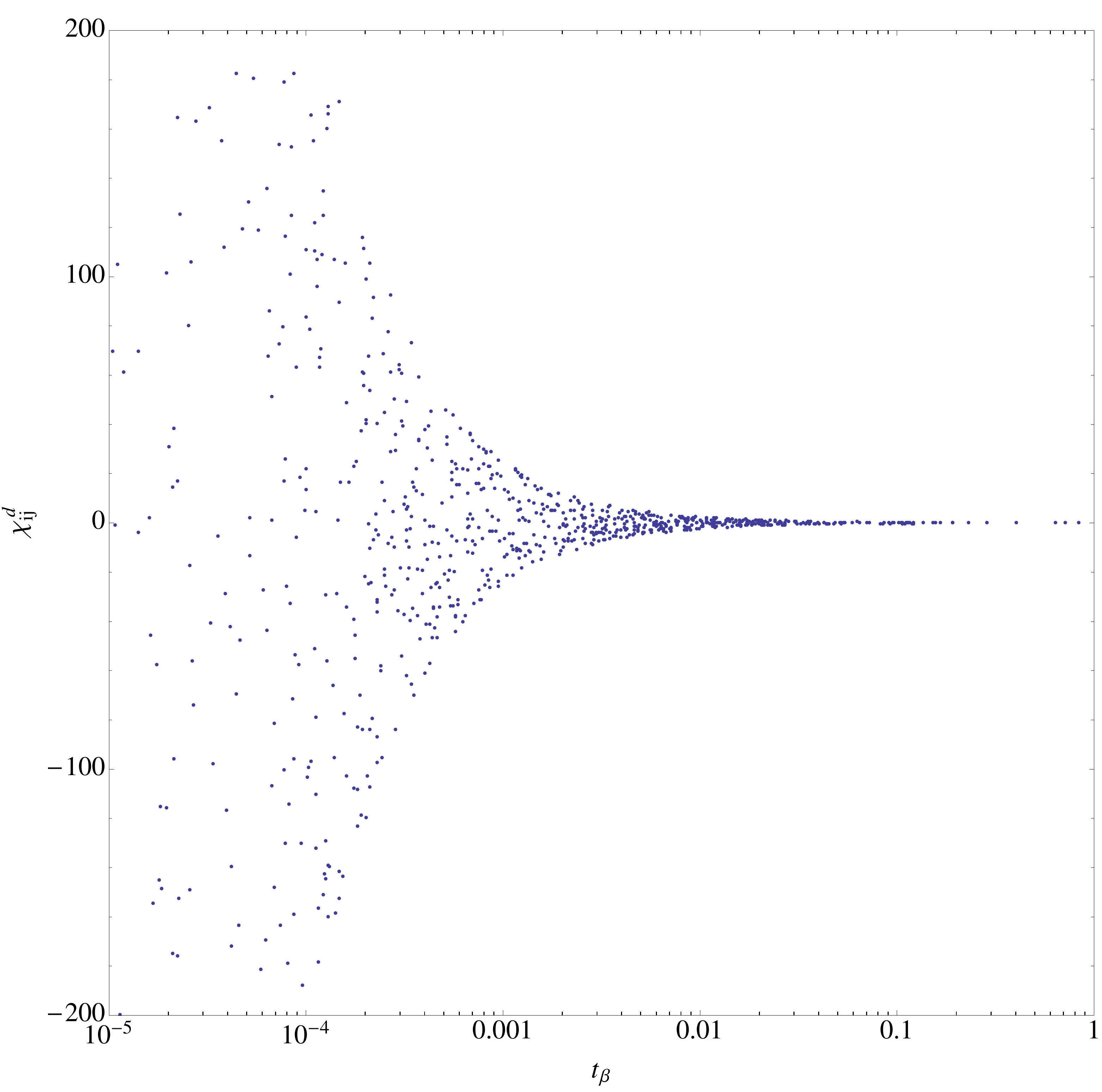}
}
\caption{Scattering plot for the flavor parameters (a) $\chi^{u}_{ij} $ and (b) $\chi^{d}_{ij} $ versus $t_\beta$,. Correlation between parameters of the model.}
\label{fig:ScPl-FC-chiuup-chiddp-tb}
\end{figure*}

Fig.\ref{fig:ScPl-FC-chiuup-chiddp-tb} 
show the parameter space for the 
u- and d-type quark versus $t_\beta^{}.$ 
We used the constraints discussed previously; 
our analysis shows that it could be possible to have 
FC if we consider the quark type separately, since 
the parameter space is wider for the u-quark type.
We found if $t_\beta\lesssim 10^{-3}_{}$ then 
$-200\lesssim\chi_{ij}^{u}\lesssim200,$
while if $t_\beta\lesssim 10^{-4}_{}$ then 
$-200\lesssim\chi_{ij}^{d}\lesssim200.$
We think is possible that the FC is generated
by quark type in multi-Higgs models.
The $\chi_{ij}^{u}-t_\beta^{}-$parameter space 
is wider (fig. \ref{fig:ScPl-FC-chiuup-chiddp-tb}) respect 
to  $\chi_{ij}^{d}-t_\beta^{}-$parameter space  (fig.\ref{fig:ScPl-FC-chiuup-chiddp-tb}). And if we consider  
$\chi_{ij}^{u} = \chi_{ij}^{d}$ around zero, 
then $t_{\beta}^{}-$parameter is more than $10^{-2}.$
%
%
%
%
Higgs particle is explored with leptons in the final states as $\tau^{}_{}\mu^{}_{};$ experimental reports have shown ${\Br}(h\to \tau^{}_{}\mu^{}_{}) <1.51\times10^{-2}_{}.$ \cite{Khachatryan:2015kon, Olive:2016xmw, Flechl:2015foa}.

\begin{figure*}[!h]
\centerline{
(a)
\includegraphics[width=2.5in]{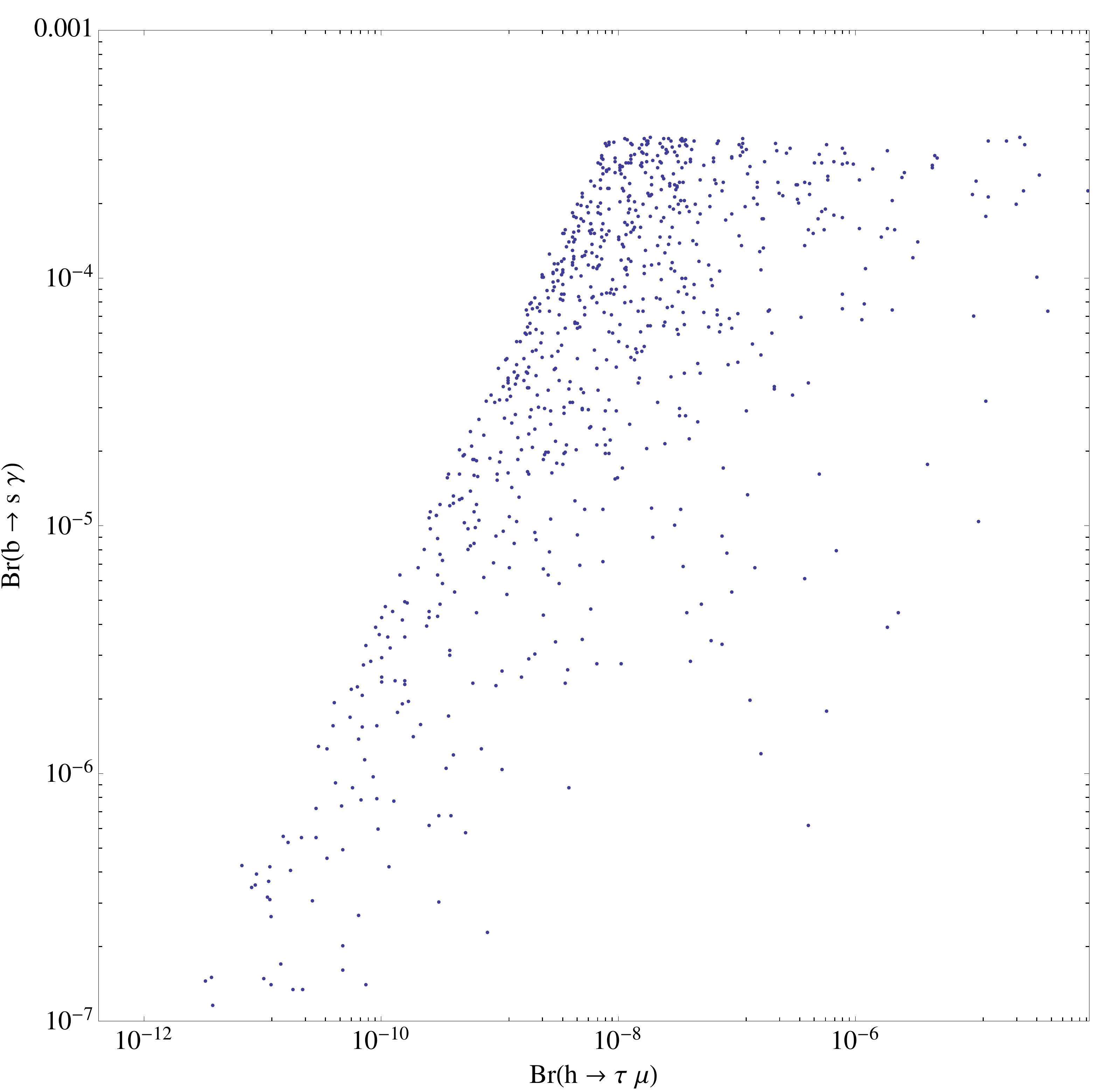}
\hfil
(b)
\includegraphics[width=2.5in]{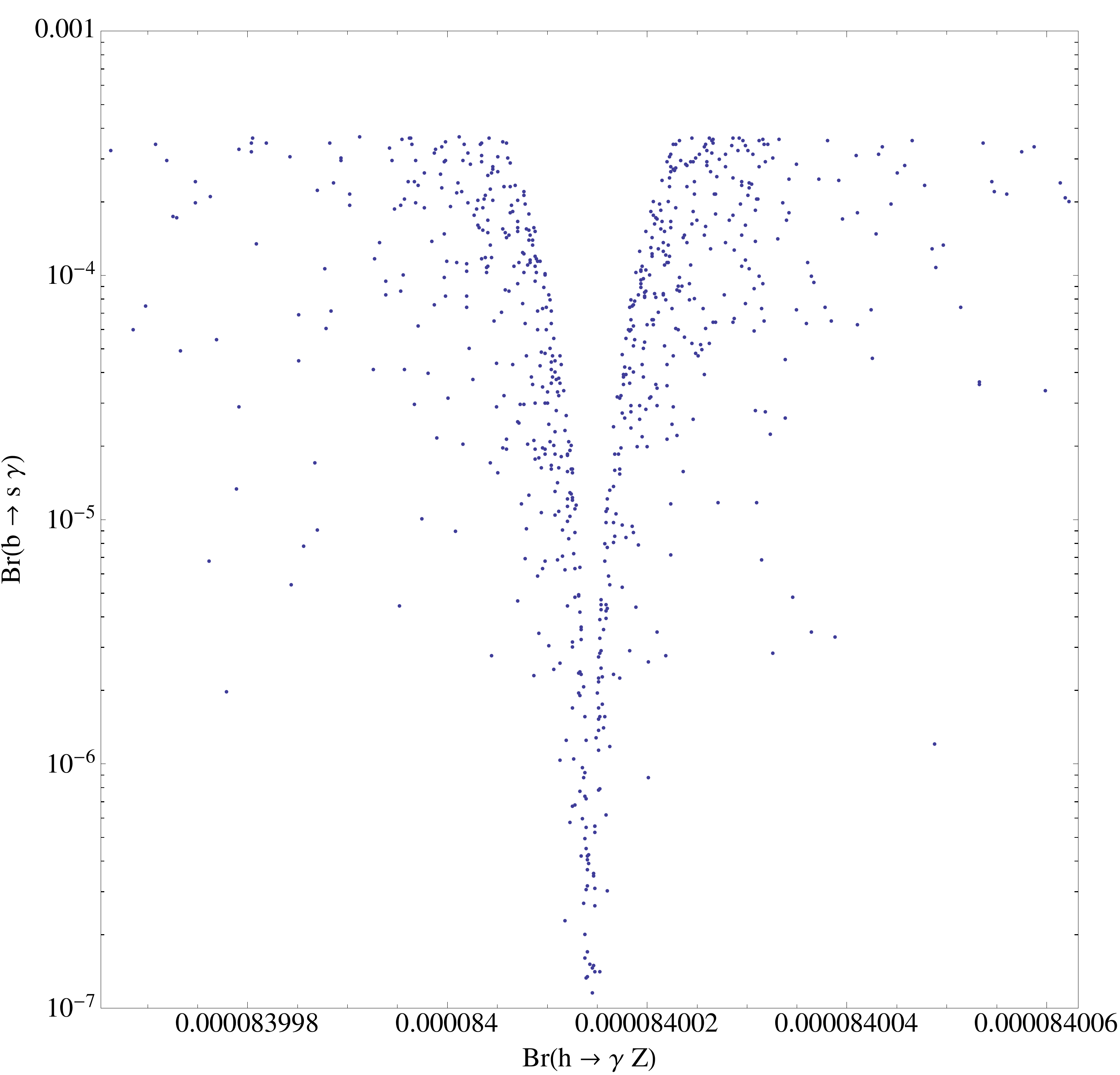}
}
\caption{\label{fig:bsg-htm-hgaZ} a) Scattering plot for the branching ratios of the processes $b\to s \gamma$ versus $h\to \tau\mu$ b) Scattering plot for the branching ratios of the processes $b\to s \gamma$ versus $h\to \gamma Z.$}
\end{figure*}

We explore the lepton processes and 
tested the relation with the d-type quarks 
as is shown in fig.\ref{fig:bsg-htm-hgaZ}.
this plot shows the correlation between $h\to \tau_{}^{} \mu_{}^{}$ 
and $b \to s\gamma$ processes. The linear and strong correlation between the $Br(b\rightarrow s \gamma)$ and $Br(h\rightarrow \tau \mu)$ is found in the regions $\sim 10^{-5}-10^{-3}$ for $Br(b\rightarrow s \gamma)$ and $\sim 10^{-9}-10^{-6}$ for $Br(h\rightarrow \tau \mu)$. Fig. \ref{fig:bsg-htm-hgaZ} shows the  region of the allowed values for the  ${\Br}(b \to s  \gamma)-{\Br}(b \to \gamma Z)$ correlation. We found a exclusion region around  ${\Br}(b \to \gamma Z) \sim 8.4002\times 10^{-5}.$ The numerical results show stability for  $Br(h\rightarrow Z \gamma)$ near to the value $8\times 10 ^{-5}$ meanwhile the $Br(b\rightarrow s \gamma)$ decreases rapidly for the same random values. The behavior of the $Br(b\rightarrow s \gamma)$ is controlled by the FC Yukawa coupling $Y^d_{23}$, which could be almost zero. By other side, the behavior of the $Br(h\rightarrow Z \gamma)$ is controlled by the dominant Feynman diagram with top quark as the internal line, which will give a proportional contribution to the Yukawa coupling without FC, $Y^u_{33}$. 
%
%
%
%

In fig. \ref{fig:bsg-htm-hgaZ}, 
we consider the constraints 
coming from  
$t\to cV, b\to s\gamma, 
h\to l_i^{} l_j^{}, h \to \gamma Z,$
to show the correlation between 
branching ratios $b\to s \gamma$ and 
$h \to \gamma Z$ and we found 
excluded regions.
The dark regions are the highly allowed as it 
is shown in figure. We generated randomly the parameter set, as follow,
$-200\leq \chi_{ij}^{u,d,l}\leq 200, 
0\leq t_\beta^{}\leq 100, 
350~{\GeV}\leq m_{H, H^\pm_{}}^{}\leq 1000~{\GeV},$ 
as well considering the 
experimental bounds for the 
$t, b$ and $h$ branching ratios 
at tree and one-loop level. 
Our calculation introduces the  
Passarino-Veltman functions implemented on 
Looptools as was mentioned above.

\begin{figure}[!h]
\centering
\includegraphics[scale = 0.4]{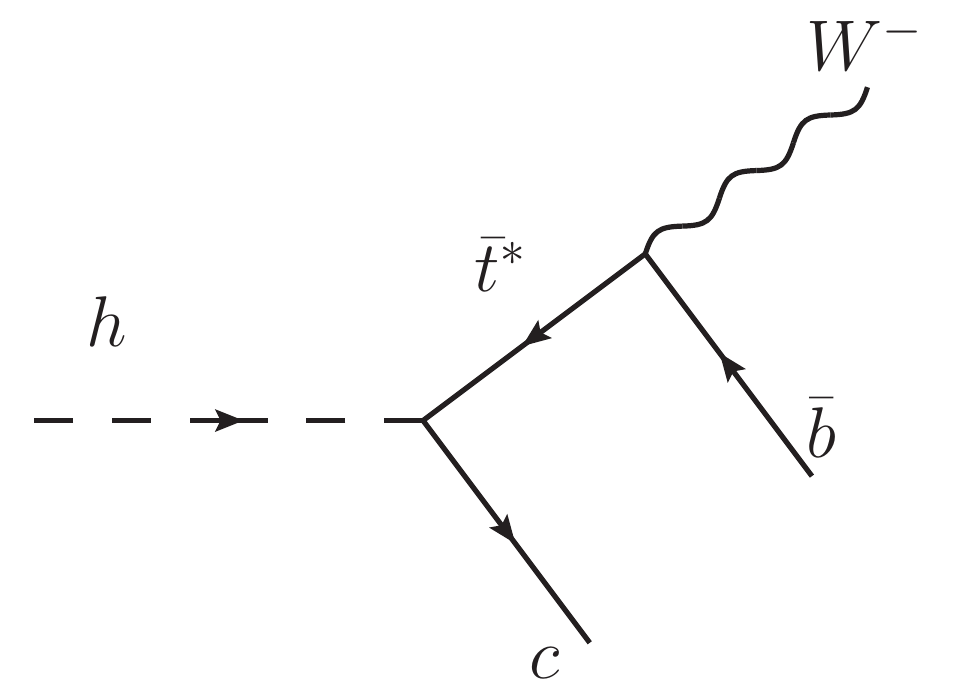}
\caption{\label{fig:FD_h-ts-c}
The Feynman diagram for the $h \to t^*_{} c.$}
\end{figure}

Fig. \ref{fig:FD_h-ts-c} shows the Feynman 
diagram for our proposal process to be explore 
considering the FC and test the standard model.
%
%
%
%

The process that is represented in 
fig. \ref{fig:FD_h-ts-c} 
was proposed by ref. \cite{DiazCruz:2012xc} to 
explore the flavor-changing modes.

\begin{figure}[!h]
\centering
\includegraphics[scale = 0.45]{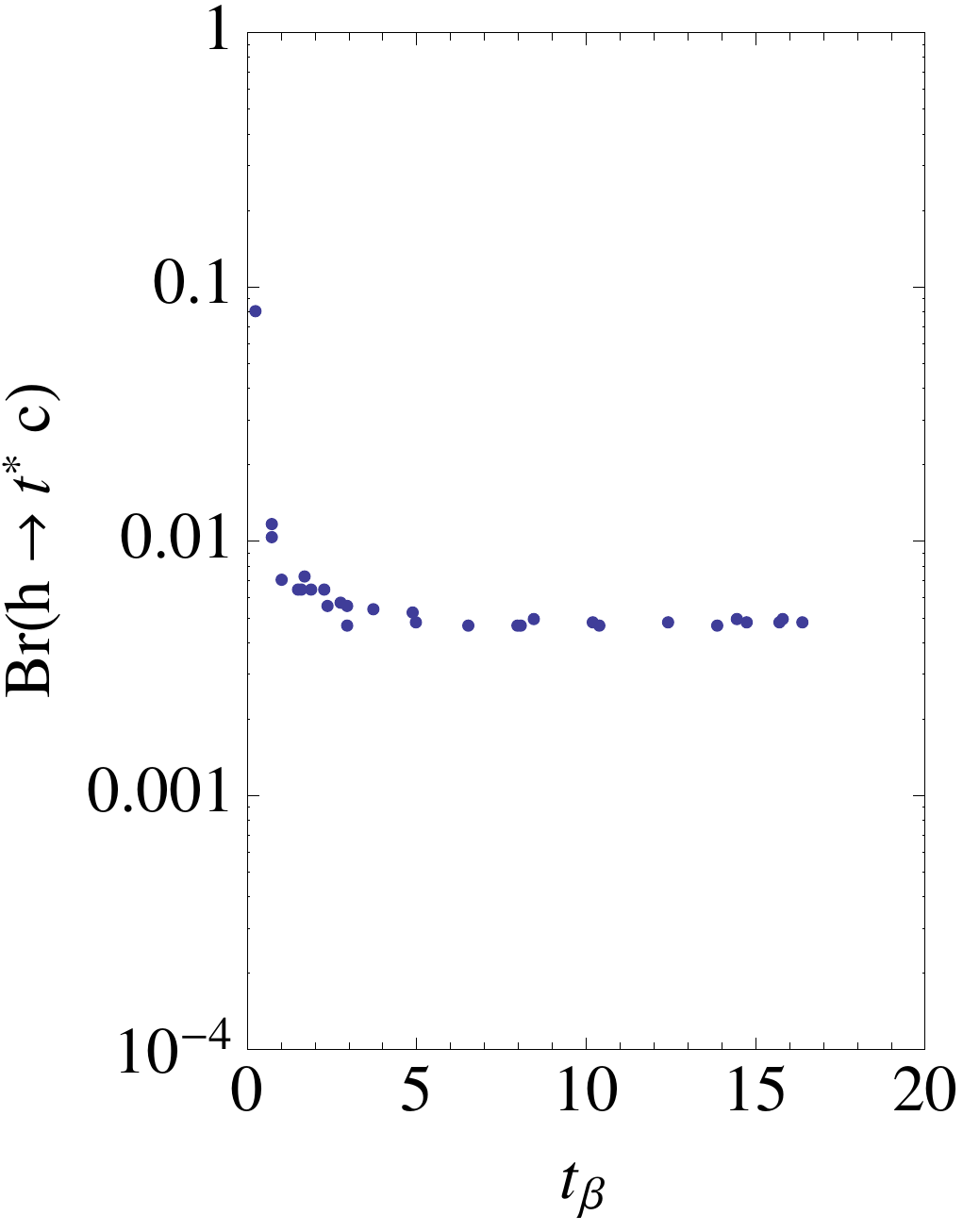}
\caption{\label{fig:Br_h-ts-c} The ${\Br}$ for the process $h\to t^{*}_{} c$ that we have predicted.}
\end{figure}

If we consider the $W^{-}$ decays to 
$\nu_l^{} l^{}_{}$ then  
we obtain ${\Br}$ of order $10^{-4},$ therefore
our results are interesting if we compare to the 
experimental results 
$\big({\Br}\sim2.45\times10^{-3}\big)$ 
as is shown in ref. \cite{Dittmaier:2011ti} 
for $m_h = 125 \GeV;$
for this channel, our results is plotted 
fig. \ref{fig:Br_h-ts-c}  versus $t_\beta^{}.$
We found ${\Br}(h \to t^{*}_{}c)\sim 10^{-2}$ for the 
$1\lesssim t_\beta^{}\lesssim20,$ 
which is a very interesting channel to explore 
in the LHC or the next generation of colliders.

%
%
%
%

Fig. \ref{fig:Br_h-ts-c} shows an interesting channel 
to probe new physics and we expect the next 
experimental results to test the THDM-III as the 
simplest SM extension. Analysis in other context 
have been made in ref. \cite{Yang:2016zyr}, differences in the results are associated to the differents models.
In ref.\cite{Belusca-Maito:2016axk} we found 
a scenario one order of magnitude above, however 
our results consider freedom in the parameters. 
Then it is possible to have $Br(h \to Wc\bar{b})\sim 10^{-3},$ 
but it fixes the other parameters as is shown in 
table \ref{ta:prmtrs-hWbc}. We note that the numerical analysis shows the $\Br(h\to t^*c)$ converges to the value $\sim 5\times10^{-3}$ when $\beta$ is large, $\tan\beta>20$. Figure \ref{fig:Br_h-ts-c} is plotted for $\tan\beta$ in $[ 0, 20 ]$, however,  for instance a random point $\tan\beta = 34.3609$  from the numerical analysis is written in table \ref{ta:prmtrs-hWbc} to show the convergence. 


\begin{table}[!h]
\caption{Representative values for the set of parameters associated to the fig.\ref{fig:Br_h-ts-c}.}
\label{ta:prmtrs-hWbc}
\centering
\begin{tabular}{|c||c||c||c||c||c|}
\hline
$\Br(h\to t^*c)$ & $t_\beta$ & $\chi_{ij}^{u}$ & $\chi_{ij}^{d}$  & $m_{H^{\pm}}^{} ~[{\GeV}]$ &$m_H^{}~[{\GeV}]$\\ 
\hline
 0.00477& 34.3609 & -0.41433 &  0.98872 & 498.982 & 981.502 \\ 
 0.00468  & 10.4314 & 1.55977 & 0.95850 &  399.117 &  982.514  \\ 
 0.00559 &  2.33911& 0.08961 &  1.11261&  510.238 &  541.895 \\
\hline
\end{tabular}
\end{table}
%
%
Table \ref{ta:prmtrs-hWbc} shows the 
allowed values for the parameters 
considering FC couplings, THDM-III, alignment 
$(\beta-\alpha=\pi/2- \delta)$ and the process 
represented in the fig.\ref{fig:FD_h-ts-c}.

\section{\label{sec:Discu-Concl} Conclusions}
We have explored the $h\to \gamma\gamma$ 
processes, and checked the high suppression 
for the FC 
in the THDM-III context  
\cite{Gaitan:2016rlq, orduz-2016:THDMIII}. 
Our exploration in the $\chi_{ij}^{u}-\chi_{ij}^{d}-$parameter space showed the allowed regions for different 
$t_\beta^{}$ values (see fig. \ref{fig:chiuu-chidd-THDM}). 

Besides
we explore the different modes for 
Higgs decays  (see fig. \ref{fig:Br0101-Br011}).
After we considered the experimental constraints 
to get scattering plots for the FC parameters and 
some relevant decay modes.
We expected the next results to figure out the 
FC and its implications in the scalar sector.
Our results showed ${\Br}(h \to \mu \tau)\lesssim10^{-5}$ 
and  ${\Br}(h \to \gamma Z)\sim 10^{-6}$ as long 
as ${\Br}(b \to s \gamma) \lesssim10^{-4}$  as 
is shown in fig. \ref{fig:bsg-htm-hgaZ}, those 
engaging values for exploring in LHC.
Besides we predicted ${\Br}(h \to t^*_{}c^{}_{})
\sim 10^{-3}$ for $1\lesssim t_\beta^{}\lesssim20$, 
this a feasible channel to explore 
our model in the LHC. 
For the on-shell top quark we obtain the 
${\Br}(h \to Wc\bar{b})\sim 4\times10^{-3}$ 
for $5\lesssim t_\beta\lesssim20$ and alignment constraint.

%
%
\section*{Acknowledgment}
This work was supported by projects PAPIIT-IN113916 in DGAPA-UNAM, PIAPI1528 in FES-Cuautitlan UNAM and Sistema Nacional de Investigadores (SNI) M\'exico. JO was supported by postdoctoral scholarships ({\it{Programa de becas posdoctorales}}) at DGAPA-UNAM. JO also would like to express a special thanks to the Mainz Institute for Theoretical Physics (MITP) for its hospitality and support. JO is grateful J. Kopp and I. Bigi, M. Carena, G. Lopez, and Roig for the discussion and comments to enrich this work.

%

\end{document}